\documentclass[twocolumn]{aastex62}
\usepackage{bm}
\usepackage{amsmath}
\usepackage{booktabs}
\usepackage{graphicx}
\usepackage{comment}

\def\ltsima{$\; \buildrel < \over \sim \;$}
\def\simlt{\lower.5ex\hbox{\ltsima}}
\def\gtsima{$\; \buildrel > \over \sim \;$}
\def\simgt{\lower.5ex\hbox{\gtsima}}

\usepackage{color}

\submitjournal{AAS Journals}

\shorttitle{1D hydro CE modelling: binary NS formation}
\shortauthors{T. Fragos, J. Andrews et al.}

\newcommand{\Msun}{\ifmmode {M_{\odot}}\else${M_{\odot}}$\fi}
\newcommand{\Rsun}{\ifmmode {R_{\odot}}\else${R_{\odot}}$\fi}
\newcommand{\mesa}{{\tt MESA}}

\begin{document}

\title{The Complete Evolution of a Neutron-Star Binary through a Common Envelope Phase Using 1D Hydrodynamic Simulations}

\correspondingauthor{T.\ Fragos}
\email{Anastasios.Fragkos@unige.ch}

\author[0000-0003-1474-1523]{Tassos Fragos}
\affiliation{Geneva Observatory, 
University of Geneva, 
Chemin des Maillettes 51, 
1290 Sauverny, Switzerland}
\affiliation{Niels Bohr Institute, University of Copenhagen, Blegdamsvej 17, 2100 Copenhagen, Denmark}

\author[0000-0001-5261-3923]{Jeff J.\ Andrews}
\affiliation{Niels Bohr Institute, 
University of Copenhagen, 
Blegdamsvej 17, 2100 Copenhagen, Denmark}
\affiliation{Foundation for Research and Technology-Hellas, 
100 Nikolaou Plastira St., 
71110 Heraklion, Crete, Greece}
\affiliation{Physics Department \& Institute of Theoretical \& Computational Physics, 
P.O Box 2208, 
71003 Heraklion, Crete, Greece}

\author[0000-0003-2558-3102]{Enrico Ramirez-Ruiz}
\affiliation{Niels Bohr Institute, 
University of Copenhagen, 
Blegdamsvej 17, 2100 Copenhagen, Denmark}
\affiliation{Department of Astronomy \& Astrophysics, University of California, Santa Cruz, CA 95064, USA}

\author[0000-0001-6181-1323]{Georges Meynet}
\affiliation{Geneva Observatory, 
University of Geneva, 
Chemin des Maillettes 51, 
1290 Sauverny, Switzerland}

\author[0000-0001-9236-5469]{Vicky Kalogera}
\affiliation{Department of Physics and Astronomy, Northwestern University, 2145 Sheridan Road, Evanston, IL 60208, USA}
\affiliation{Center for Interdisciplinary Exploration and Research in Astrophysics (CIERA), Northwestern University, 2145 Sheridan Road,
Evanston, IL 60208, USA}
\affiliation{CIFAR Fellow}

\author[0000-0001-8805-2865]{Ronald E.\ Taam}
\affiliation{Department of Physics and Astronomy, Northwestern University, 2145 Sheridan Road, Evanston, IL 60208, USA}
\affiliation{Center for Interdisciplinary Exploration and Research in Astrophysics (CIERA), Northwestern University, 2145 Sheridan Road,
Evanston, IL 60208, USA}

\author[0000-0001-8952-676X]{Andreas Zezas}
\affiliation{Foundation for Research and Technology-Hellas, 
100 Nikolaou Plastira St., 
71110 Heraklion, Crete, Greece}
\affiliation{Physics Department \& Institute of Theoretical \& Computational Physics, 
P.O Box 2208, 
71003 Heraklion, Crete, Greece}

\begin{abstract}
Over forty years of research suggests that the common envelope phase, in which an evolved star engulfs its companion upon expansion, is the critical evolutionary stage forming short-period, compact-object binary systems, such as coalescing double compact objects, X-ray binaries, and cataclysmic variables. In this work, we adapt the one-dimensional hydrodynamic stellar evolution code, \mesa, to model the inspiral of a 1.4\Msun\ neutron star (NS) inside the envelope of a 12\Msun\ red supergiant star. We self-consistently calculate the drag force experienced  by the NS as well as the back-reaction onto the expanding envelope as the NS spirals in. Nearly all of the hydrogen envelope escapes, expanding to large radii ($\sim$10$^2$ AU) where it forms an optically thick envelope with temperatures low enough that dust formation occurs. We simulate the NS orbit until only 0.8\Msun\ of the hydrogen envelope remains around the giant star's core. Our results suggest that the inspiral will continue until another $\approx$0.3\Msun\ are removed, at which point the remaining envelope will retract. Upon separation, a phase of dynamically stable mass transfer onto the NS accretor is likely to ensue, which may be observable as an ultraluminous X-ray source. The resulting binary, comprised of a detached 2.6\Msun\ helium-star and a NS with a separation of 3.3-5.7\Rsun, is expected to evolve into a merging double neutron-star, analogous to those recently detected by LIGO/Virgo. For our chosen combination of binary parameters, our estimated final separation (including the phase of stable mass transfer) suggests a very high $\alpha_{\rm CE}$-equivalent efficiency of $\approx$5.
\end{abstract}


\section{Introduction}
The recent detection of gravitational waves from coalescing binary black holes and most recently of a binary neutron star (NS) merger \citep[][and references therein]{2018arXiv181112907T} and the associated gamma-ray burst and kilo-nova explosions \citep{2017ApJ...848L..12A}, sparked a renewed interest in the formation of compact-object binaries. For two NSs to merge within a Hubble time, they must have an orbital separation of $\lesssim10$\Rsun, yet on their way to NS formation, all massive stars go through a supergiant phase in which they expand to radii of $\sim500-1000$\Rsun, two orders of magnitude larger than their ultimate orbital size. The common envelope (CE) phase, originally discussed in the context of cataclysmic variable formation \citep{1976IAUS...73...75P}, has been widely adopted by the binary evolution community as the mechanism responsible for forming short-period, compact-object binaries \citep[e.g.,][]{1995MNRAS.272..800H,Belczynski:2016kc}. For a modern description (and history) of the CE, we refer the reader to the thorough review by \citet{Ivanova:2013co}. 

The CE phase typically occurs when the companion of a giant star is engulfed by the giant's envelope, which now fills the combined binary potential, surrounding both the accretor and the giant star's core. The frictional torque of the companion orbiting within the fluid of the giant's envelope transforms orbital energy and angular momentum into heat and spin angular momentum of the CE, dramatically shrinking the orbit. The frictional heat dumped in the envelope causes its expansion and -- if there is enough orbital energy available -- its eventual expulsion from the system, leaving behind the core of the giant star and its companion in a close orbit. 

Quickly after its conception, the CE was studied using numerical hydrodynamics in 1D by \citet{1978ApJ...222..269T,1979A&A....78..167M,1979ApL....20...29T}. These authors showed some promising initial results; however it was quickly realized that the resulting evolution was not spherically symmetric \citep{Bodenheimer:1984fs}, which led to the abandonment of 1D simulations for over 15 years in favor of multi-dimensional simulations to account for the non-axisymmetric geometry of the problem and the turbulent processes involved. Modern simulations employ adaptive mesh refinement techniques \citep{Ricker:2008kc,TR2010,Ricker:2012gu,Passy:2012jo}, moving mesh simulations \citep{Ohlmann:2016kj}, or smooth particle hydrodynamics \citep{2015MNRAS.450L..39N,2016MNRAS.455.4351P,Pejcha:2016kf}. These studies highlight the difficulties of modeling the CE evolution in 3D, as they typically require several million CPU-hours per simulation and still may not include all the necessary physics such as energy and radiation transport and a realistic equation of state. Even without all the necessary physics required, efforts to model this process are hampered by the wide dynamic range in both temporal and spatial scales \citep{1994ApJ...422..729T,1996ApJ...471..366R,2000ARA&A..38..113T}. 

Building off previous models by \citet{1978ApJ...222..269T}, \citet{1979A&A....78..167M}, \citet{Podsiadlowski2001}, and \citet{Ivanova:thesis} \citep[and more recent work by][]{2016MNRAS.462..362I,2017MNRAS.470.1788C}, we present one-dimensional simulations of the evolution of a binary through the CE phase, from the onset of the dynamically unstable mass-transfer until the successful ejection of the CE. In this first work, we present the methodology of our approach and describe in detail one test case simulation, relevant to the formation of a coalescing binary NS. This simulation is performed with a modified version of the stellar evolution code \mesa\ \citep{Paxton2011,2013ApJS..208....4P,2015ApJS..220...15P,2018ApJS..234...34P}, taking advantage of the accurate energy transport and detailed microphysics of a stellar structure code as well as the hydrodynamic capabilities implemented in recent code releases. In section \ref{sec:method} we describe our adaptations to \mesa. We provide the results and discussion of our simulations in Section \ref{sec:results}. Finally, we provide some conclusions in Section \ref{sec:conclusions}.

\section{Method}
\label{sec:method}

In what follows we describe the setup of our simulation and the modifications we have made to the standard \mesa\ code. We use \mesa\ version 9793 and the \mesa\ software development kit version 20170802.

\subsection{Model initialization and onset of the dynamical instability}

We initialize our binary system, consisting of a 12\Msun\ zero-age main sequence star and a 1.4\Msun\ point-like companion (representing a NS) at an orbital separation of 972\,R$_{\odot}$, using the binary module within \mesa. We assume a circular orbit and neglect both wind mass loss and tidal interactions. Therefore, the orbital separation is constant until the star expands to $\approx 550\,\rm R_{\odot}$ as a giant star, overfilling its Roche lobe. The rate of mass transfer due to Roche lobe overflow is calculated using the implicit numerical scheme formulated by \citet{1990A&A...236..385K}, which allows for a potentially significant overflow of the Roche lobe by the donor star. Mass transfer is assumed to be conservative up to the Eddington limit, with the excess material being ejected with the specific orbital angular momentum of the NS. 

We continue our simulation until the donor star overfills either of the outer Lagrangian points $L_{\rm 2}$ or $L_{\rm 3}$ (using the fitting formulae for the equivalent $R_{L_{\rm 2}}$ and $R_{L_{\rm 3}}$ radii by Misra et al., in preparation), at which point we consider that the binary enters dynamically unstable mass transfer.  The donor's mass loss rate at the time of $L_{\rm 2}$/$L_{\rm 3}$ overflow onset is $\approx 10^{-2}\rm\, M_{\odot}\, yr^{-1}$, and a few percent of the donor's mass has already been lost.

\subsection{Hydrodynamic modelling of the CE phase}

After we resolve the onset of dynamical instability, we switch to the CE \mesa\ module that we developed specifically for this work. Throughout this stage of simulation, we use an energy-conserving implicit hydrodynamic solver which has improved performance when simulating rapid variations (such as shocks) outside the hydrostatic regime \citep[for details see section 4 of][]{2015ApJS..220...15P}. When using this numerical scheme, the outer boundary conditions are not set by standard stellar atmosphere models, which in any case are inadequate for the problem at hand due to supersonic outflow velocities and very low gas densities developed in the outer part of the CE during the simulation. Instead, to close the stellar structure equations at the outer boundary of the star, we ensure that the compression vanishes ($d\rho/dm=0$) and the temperature is equal to that of a black body for the given luminosity and radius of the shell, and we arbitrarily set the optical depth of the outer shell to $10^{-4}$. In our calculations, we also include corrections to the energy equation due to rotation, although the current implementation of the implicit scheme does not allow analogous rotation corrections to the momentum equation. During our code testing, we encountered numerical instabilities when simulating stellar envelopes of sufficiently low density and temperature which resulted from a transition between the OPAL and the HELM equations of state. To handle this, we use numerical partial derivatives of the internal energy, entropy and gas pressure, instead of the interpolated values provided by the equation of state tables. In the version of \mesa\ used in our simulations this feature was in an experimental stage, but since then, it has been further refined and is now a standard feature in the most recent \mesa\ versions. 

Using the above adaptations, we initialize the second phase of our simulation by placing the NS just below the surface of the envelope of the donor star, at a radius equal to 99\% of the total donor's radius. We further set the donor star to have a uniform rotation equal to 95\% of the orbital angular frequency. This amounts to the assumption that during the expansion phase of the donor star and the onset of the mass transfer, there was enough time to, almost, synchronize the rotation of the donor's envelope with the orbit.

We consider the NS to be a point mass moving in a circular orbit with a Keplerian velocity within the envelope of the giant. The NS feels a gravitational drag force causing it to spiral deeper within the giant; it also feels a hydrodynamic drag force and tidal forces, however both 3D hydrodynamic simulations and order of magnitude estimates indicate the gravitational drag force dominates \citep[e.g.][]{1978ApJ...222..269T,Passy:2012jo}. We add the dissipated energy, determined by the gravitational drag, as an extra heating term within the star's envelope which is deposited within roughly one accretion radius, $R_A$, of the compact object. Each shell within the accretion radius is weighed by a factor $\exp{\left[-(\Delta r/R_A)^2\right]}$, where $\Delta r$ is the radial distance of each mass shell from the NS orbit. The drag force is calculated using the fitting formula by \citet{2015ApJ...803...41M} in the supersonic regime, which provides corrections to the standard expression due to the density gradient in the structure of the star, and by analytic estimates from \citet{Ostriker:1999jz} in the subsonic regime, with the two being smoothly blended for Mach numbers between 0.9 and 1.1. We additionally include accretion luminosity as an energy source, calculated from the Bondi rate but limited to the Eddington rate. This provides a reasonable lower limit to the accretion luminosity as the flow is unable to get to the neutrino cooling regime \citep{2015ApJ...798L..19M}. These simulations find accretion rates somewhat larger than the Eddington-limited rate, but nevertheless only total to $\lesssim$0.1~\Msun. 

At every time step we update the orbital energy based on the gravitational drag, from which we calculate the NS's position, assuming a circular orbit. Although we conserve energy, since we enforce orbital circularity, we cannot simultaneously conserve angular momentum. In future revisions of our simulations we plan to relax the assumption of a circular binary orbit, which will allow us to also self-consistently spin up the envelope during the inspiral. Here, we still follow the angular momentum transport within the envelope as it restructures, but we do not apply any torque to it due to the NS's inspiral.

As an additional caveat, we note that we ignore the back-reaction of the NS's gravitational pull on the giant star's envelope. Properly accounting for this effect requires multi-dimensional simulations.

Our models do not formally include mass loss, and we track the evolution of the outer envelope even as it expands to radii beyond 10$^4$ \Rsun. This allows us to analyze the long-term behavior of the envelope and determine whether it remains truly unbound or whether it goes through oscillation cycles as recently studied by \citet{2017MNRAS.470.1788C}.  Our simulation is terminated due to numerical reasons when the outer layers of the expanding envelope reach temperatures and densities near the edges of our tabulated equation of state.

\section{Results and Discussion}
\label{sec:results}

\subsection{Phenomenological Description}

\begin{figure*}
    \centering
    \includegraphics[width=0.48\textwidth]{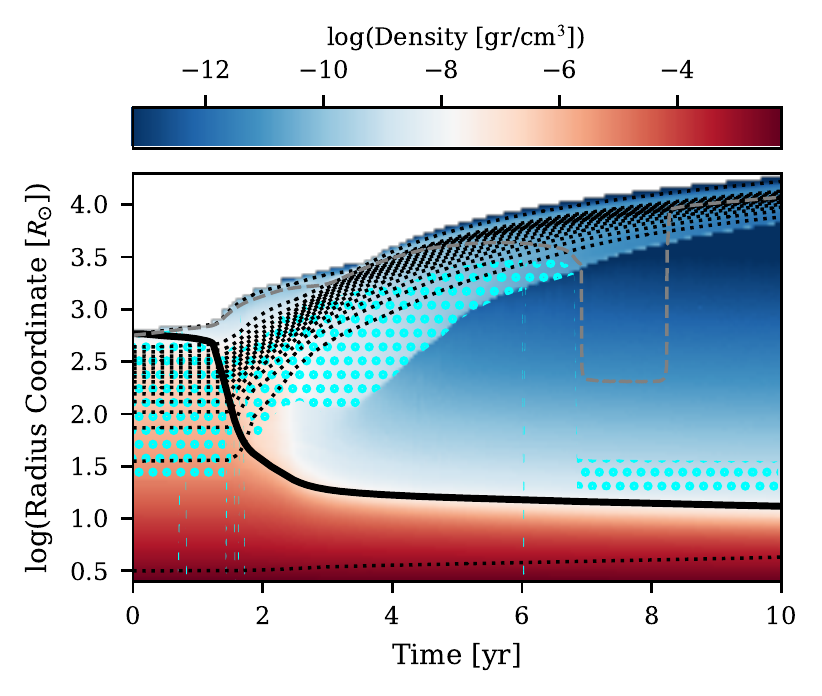}
    \includegraphics[width=0.48\textwidth]{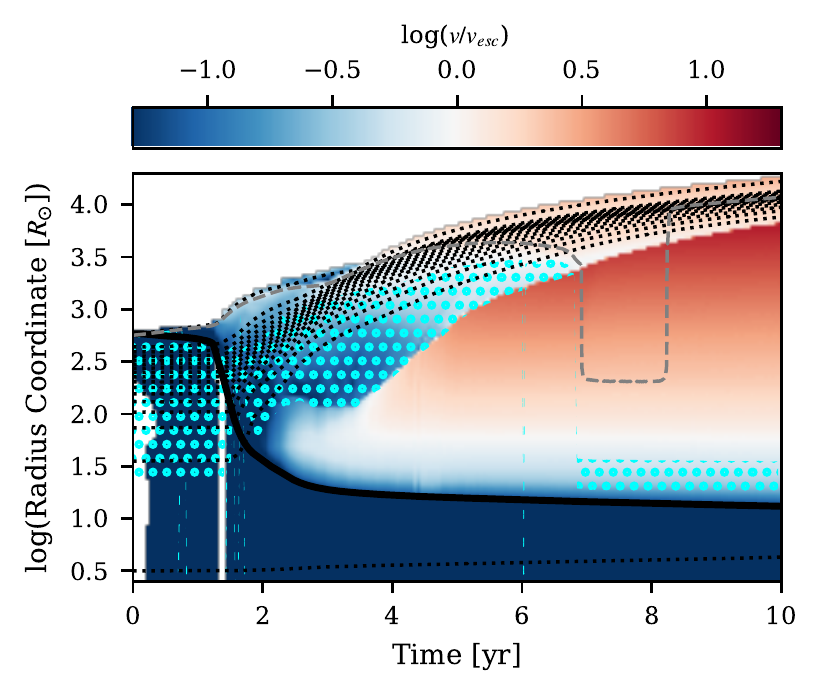}
    \includegraphics[width=0.48\textwidth]{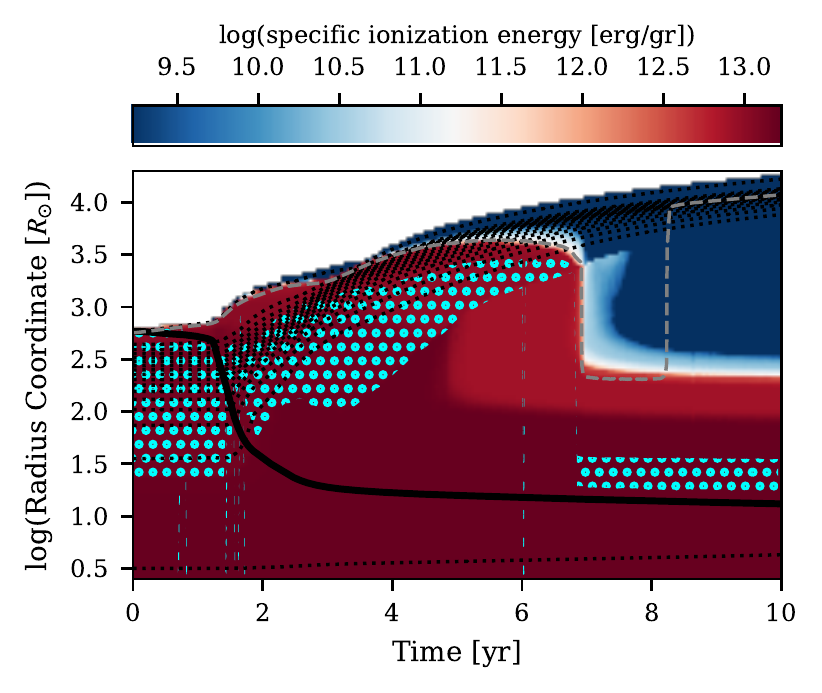}
    \includegraphics[width=0.48\textwidth]{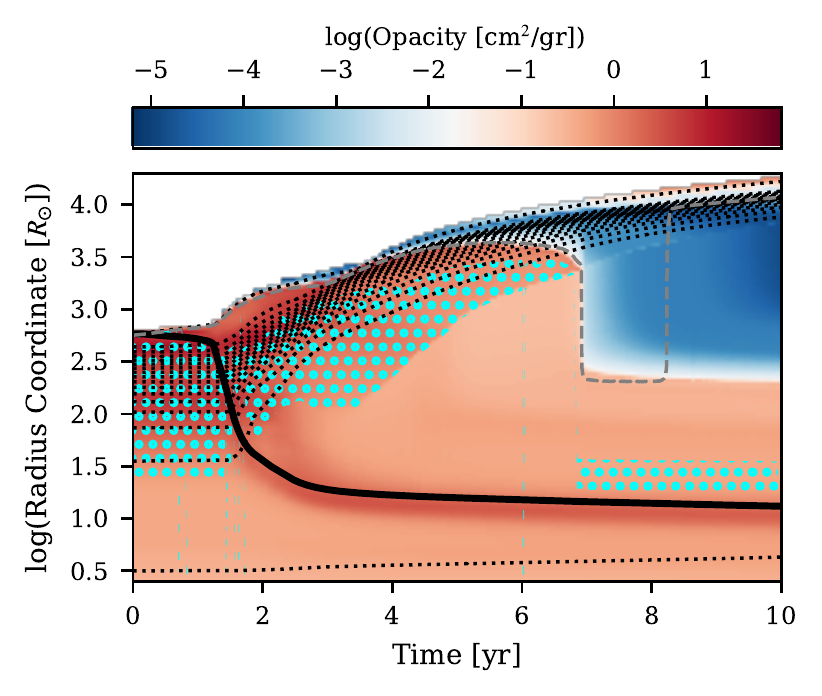}
    \caption{Kippenhan-type diagrams showing the evolution of the 12 \Msun\ giant star's envelope structure (the star's core is not included for clarity), in terms of density (top left), radial velocity (top right), ionization energy (bottom left), and opacity (bottom right), for the first 10 years of the CE phase. The position of the NS's orbit is denoted by the thick, black line. Dotted black lines denote surfaces of constant mass (spaced at $0.5\,\rm M_{\odot}$ from each other), while the dashed gray lines indicate the position of optical depth of 10. Areas highlighted by cyan "circles" denote convective regions.
    }
    \label{fig:kippenhahn}
\end{figure*}

The results of our simulation of a 1.4 \Msun\ NS inspiraling within the envelope of a 12 \Msun\ giant star are shown in Figure \ref{fig:kippenhahn}. The inspiral shows the three distinct phases of CE evolution described in \citet{Podsiadlowski2001} and \citet{Ivanova:2013co}. The initial phase when the NS's position is roughly constant shows the ``loss of co-rotation". Throughout this phase, the envelope is nearly co-rotating with the engulfed NS, and therefore the weak gravitational drag on the NS leads to only a gradual decay in the NS orbit. Since the thermal timescale from the position of the NS to the surface of the CE is extremely short ($10^{-4}-10^{-2}\,\rm yr$), much of the injected energy quickly radiates away, leaving the envelope relatively unaffected by the presence of the NS.

After $\sim$1.5 years of slow inspiral, when co-rotation is broken,  the system enters the ``plunge-in" phase in which the NS's orbit rapidly shrinks from several hundreds of \Rsun\ to $\sim$20 \Rsun. Concurrent with the shrinking of the NS's orbit, the giant star's envelope shows rapid expansion to  $\sim$2000 \Rsun. The expansion of the whole envelope can be seen by following the time evolution of the thin, black, dotted lines in Figure \ref{fig:kippenhahn} which denote surfaces of constant mass. After the dynamical ``plunge-in" phase, the inspiral decelerates and  almost stalls during the ``self-regulated regime", which we model for $\simeq$18 years. For clarity, Figure \ref{fig:kippenhahn} shows only the first 10 years of the simulation.

The top left panel of Figure~\ref{fig:kippenhahn} shows the evolution of the density profile of the donor star. The effect of the NS's inspiral on the envelope structure is profound; at a radius of 100 \Rsun, the density drops by roughly four orders of magnitude, from $\sim $10$^{-6}$ g cm$^{-3}$ at the start of the simulation to $\sim $10$^{-10}$ g cm$^{-3}$ after 10 years.  By the end of the simulation, the majority of the envelope's mass is found in an extended, low density envelope at very large radii ($\sim $10$^4$\Rsun),  with temperatures low enough ($\sim 600\,\rm K$) that dust formation can occur. This is the cause of the increased opacity at the outer layers and the rise of optical depth 10 line closer to the surface after $\simeq$8\,yr, as seen in the bottom right panel of Figure~\ref{fig:kippenhahn}.  In addition, it is worth noting that throughout the self-regulated regime, the NS is found at an almost constant density of $\sim $10$^{-8}$ g cm$^{-3}$, just outside a region with a very steep density gradient.

The top right panel of Figure \ref{fig:kippenhahn} shows the radial velocity profile of the envelope, scaled to the local escape velocity. After the first $\simeq$4\,yr, the envelope expansion occurs at a nearly constant velocity of $\simeq$20 km/s. White indicates material leaving at the escape velocity, with red showing material that escapes faster. By the end of the simulation, all the material at large radii ($>$8.5 \Msun\ of the envelope) has a velocity above the escape velocity, indicating the envelope has been successfully ejected.

The bottom left panel of Figure \ref{fig:kippenhahn} shows the available ionization energy of the envelope. At a given time, the sharp transition in radius from red to white/blue denotes the hydrogen recombination front; the majority of the envelope has recombined by the end of the simulation. At the same time, a helium recombination front exists, which can be seen by the transition between two shades of red, at radii of $\sim 10^{2.5}$ \Rsun\ initially and $\sim 10^{2}$ \Rsun\ after $\simeq 5\,\rm yr$. The latter occurs at high enough opacities (shown in the bottom right panel of Figure \ref{fig:kippenhahn}) that photons can be reabsorbed by escaping material before they expand to large enough radii, cool and recombine hydrogen as well. There has been much discussion in recent literature addressing the question of whether the radiation released by recombination will be absorbed by the envelope and aid in its ejection \citep[e.g.,][]{2018ApJ...858L..24I} or whether the radiation will be quickly transported to the outer layers of the star where it can be radiated away \citep[e.g.,][]{2018MNRAS.478.1818G}. However, here we note that, as \citet{2016A&A...596A..58K} have also demonstrated, in massive stars, the total amount of energy released by recombination adds up to only a small fraction ($\lesssim 10\%$ in our case) of the energy required to eject the envelope.

\subsection{Energetics of the inspiral}

\begin{figure}
    \centering
    \includegraphics[width=0.99\columnwidth]{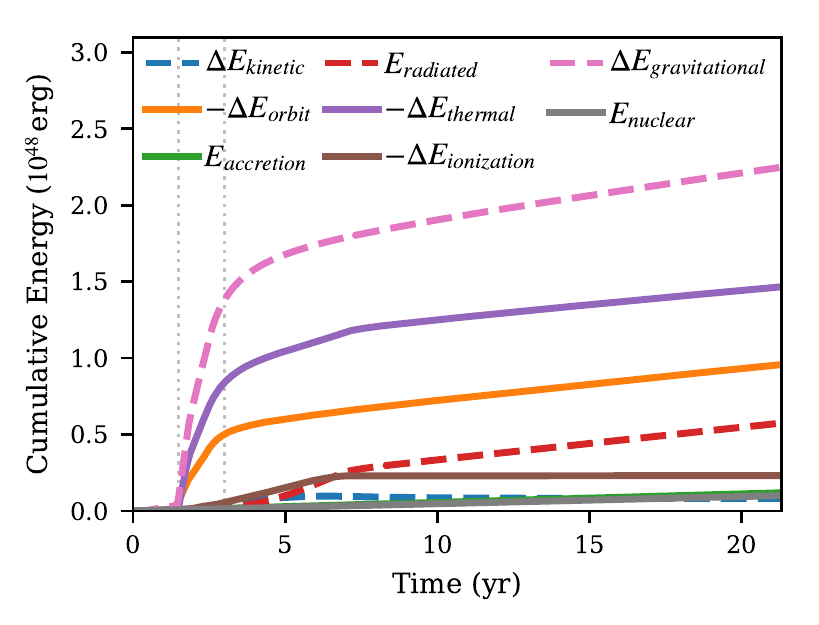}
    \caption{Cumulative energy sources (solid lines) and sinks (dashed lines), calculated by integrating the entire donor star structure, as a function of time. The two thin, dotted vertical lines denote the approximate transition times between the phases of ``loss of co-rotation", ``dynamical plunge-in" and ``self-regulated inspiral".}
    \label{fig:cum_energies}

\end{figure}

To evaluate the importance of the various physical processes, we calculate each of their energies throughout the simulation. Figure~\ref{fig:cum_energies} shows the cumulative energy budget over the entire star as a function of time for each relevant source (solid lines) and sink (dashed lines). Specifically we show the change in potential/gravitational, thermal (which includes energy stored in radiation), ionization and kinetic energy of the star. Furthermore, we track the cumulative energy injected in the envelope due to the gravitational drag onto the NS that converts orbital energy into heat ($\Delta E_{orbit}$) as well as accretion onto the NS ($E_{accretion}$), limited to the Eddington rate. Note that while Figure \ref{fig:cum_energies} shows that accretion is not currently an important energy source, if the rate were super-Eddington by a factor of 10 or more, accretion could play a significant role \citep[the reader is referred to][for a detailed discussion of accretion onto a NS within a common envelope]{2015ApJ...798L..19M}. We plan to explore the effect of super-Eddington accretion in future work. Finally, we track the total energy that is radiated away from the surface of the star and the nuclear energy released. We have checked that the sum of all energy sources and sinks equals approximately zero throughout the evolution, ensuring energy conservation to $\sim 1\%$.

During the first 1.5\,yr of the CE evolution, i.e. phase of ``loss of co-rotation", there is little evolution in the energies shown in Figure \ref{fig:cum_energies} since the envelope structure remains relatively unchanged. However, during the ``plunge-in" phase (between $\simeq$1.5\,yr and $\simeq$3\,yr) the envelope expands quickly, and its gravitational binding energy increases (becomes less negative) by $\simeq 1.38\times10^{48}\,\rm erg$, a factor of 2.7 larger than the orbital energy released from the inspiraling NS ($\simeq 0.51\times10^{48}\,\rm erg$). The excess energy resides in the initial thermal energy content stored in the star, which as the envelope expands, does work and, as shown in Figure \ref{fig:cum_energies}, releases an additional $\approx 0.84\times10^{48}\,\rm erg$. Other energy sources and sinks can be ignored, as they add up to $\lesssim 0.05\times10^{48}\,\rm erg$. The second dotted line in  Figure \ref{fig:cum_energies} shows the end of the “dynamical plunge-in” phase. The position of the NS within the giant's envelope where this transition occurs can be roughly estimated, directly from the initial stellar model, by finding the point where the gravitational binding energy of the envelope and its internal energy (integrated from the surface to that point) balances the change in the NS's orbital energy.

This implies that for every erg released by the decay of the NS orbit, an additional 1.7 ergs are tapped from the thermal energy content of the envelope. According to the virial theorem, the thermal energy of the envelope relates to  its gravitational energy as $E_{thermal}=-[1/(3\gamma -3)]E_{grav}$, where $\gamma$ is the adiabatic index of the envelope that is associated  to the equation of state. During the inspiral, if we inject a given amount of energy, in our case $\Delta E_{orbit}$, to the total energy of the system ($E_{thermal} + E_{grav}$), then its thermal energy will decrease by $\Delta E_{thermal}=-[1/(4-3\gamma)]\Delta E_{orbit}$ and its gravitational energy will increase by  $\Delta E_{grav} = [(3-3\gamma)/(4-3\gamma)]\Delta E_{orbit}$. 

For envelopes of massive stars, $\gamma$ lies between the values of 4/3 and 5/3, corresponding to a purely radiative gas and a monoatomic, ideal gas, respectively. In our specific model, the outer 9\,M$_\odot$ of the star initially has $\gamma$ ranging from 1.4 to 1.5, which would translate, based on the virial theorem, to $\Delta E_{grav}/\Delta E_{orbit}\simeq3$, close to the value of 2.7 that we find (Figure \ref{fig:cum_energies}). 

\begin{figure}
    \centering
    \includegraphics[width=0.99\columnwidth]{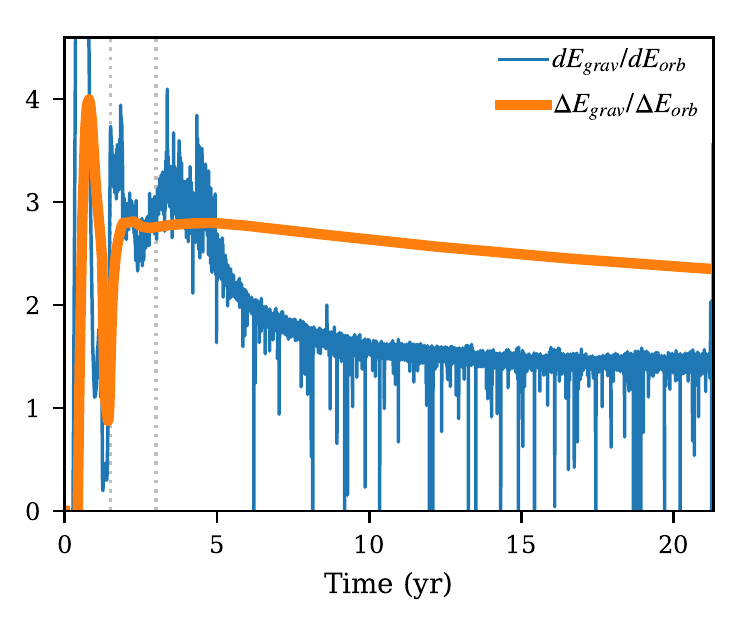}
    \caption{The instantaneous (blue) and time-averaged (orange) ratio of the envelope's gravitational energy change to the NS's orbital energy change. As in Figure~\ref{fig:cum_energies}, the two thin, dotted vertical lines denote the transition between the different phases of CE evolution.}
    \label{fig:efficiency}
\end{figure}

In Figure \ref{fig:efficiency} we show the ratio of the total change, since the beginning of the simulation, of the envelope's gravitational energy to the NS's orbital energy ($\Delta E_{grav}/\Delta E_{orbit}$; orange line),  as well as the instantaneous ratio of the changes of the two energies as a function of time ($d E_{grav}/d E_{orbit}$; blue line). Supporting our expectation from applying the virial theorem to the unperturbed model of our star, during the dynamical plunge-in and up to $\approx5\,\rm yr$ of the simulation, when radiative losses are still negligible, Figure \ref{fig:efficiency} shows that both $\Delta E_{grav}/\Delta E_{orbit}$ and $d E_{grav}/d E_{orbit}$ have average values of $\simeq$3. We note that \citet{Ohlmann:2016kj}, find a similar ratio of $\Delta E_{grav}/\Delta E_{orbit}\simeq 2.5$ in their 3D hydrodynamic simulation of a CE, albeit using a very different mass range and simulation set-up.

After 6.5\,yr, most of the giant star's envelope has recombined, releasing $0.23\times 10^{48}\,\rm erg$ of energy into the CE. This recombination reduces the envelope's opacity by 1-2 orders of magnitude, which in turn reduces the energy transport timescale from the NS to the surface of the CE by two orders of magnitude to $\approx 10^7\,\rm s$. At the same time, the inspiral timescale slows to $\approx 10^9\,\rm s$. Radiative losses become considerable, and the evolution is no longer adiabatic. Figure \ref{fig:cum_energies} shows that by the end of the simulation $0.56\times 10^{48}\,\rm erg$ have been lost from the system via radiation. 

While the injection of orbital energy still causes a commensurate release of thermal energy, radiative losses restrict that energy from being entirely used to expel the envelope. Furthermore, as the envelope expands and cools, it becomes less radiation-dominated, leading to an increase in its adiabatic index, $\gamma$. The result of both effects is a reduction in the instantaneous $d E_{grav}/d E_{orbit}$ - and therefore a gradual decrease in the time averaged $\Delta E_{grav}/\Delta E_{orb}$ - seen after $\approx$5 years in Figure \ref{fig:alpha_prediction}. One should also keep in mind that at these longer timescales, the energy added to the envelope due to accretion onto the NS and nuclear reactions from the hydrogen-burning shell underneath become non-negligible.

At the end of our hydrodynamic simulation a non-negligible hydrogen envelope still remains around the helium core. This remaining hydrogen-rich layer between the NS's orbit and the helium core, although small in mass, dominates the overall gravitational binding energy of the envelope due to its small radius. Therefore, exactly how much hydrogen remains around the helium core when the binary exits the CE phase greatly affects its final orbital separation \citep{Ivanova:2013co}. In the following section, we discuss how we extrapolate our simulation and estimate the final orbital separation and the resulting value of the common envelope efficiency parameter, $\alpha_{CE}$.

\subsection{Exiting the Common Envelope and Final Outcome}
\label{sec:exit}

\begin{figure*}
    \centering
    \includegraphics[width=0.99\columnwidth]{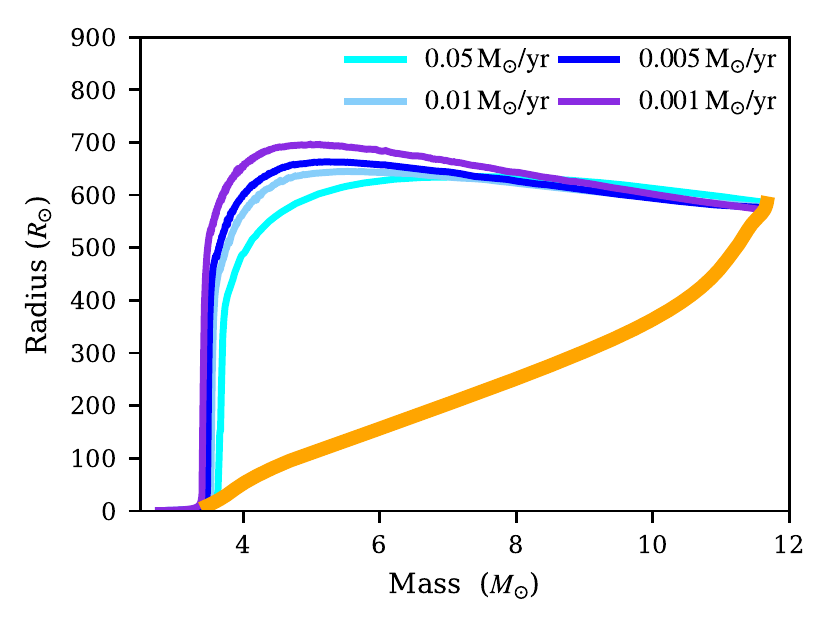}
    \includegraphics[width=0.99\columnwidth]{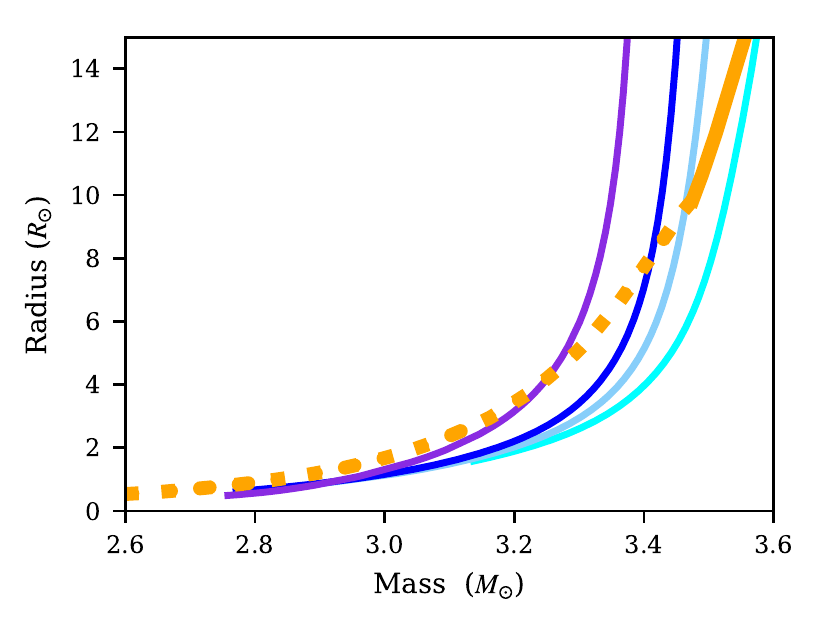}
    \caption{The solid, orange line shows the position of the NS in mass and radius coordinate of the CE as it inspirals (from right to left), based on our hydrodynamic simulation, while the cyan, light blue, purple, and magenta lines show the evolution of mass-radius relation for a 12\,M$_\odot$ giant star with initial radius of 550\,R$_\odot$ that experiences constant rapid mass loss of 0.001, 0.005, 0.01 and 0.05 M$_\odot$/yr respectively. The \textbf{right panel} is a zoom-in of the \textbf{left panel} for the last phase of the inspiral. The orange dashed line shows the predicted continuation of the NS's inspiral after the end of the hydrodynamic simulation. 
    }
    \label{fig:end_of_CE}
\end{figure*}

Figure~\ref{fig:efficiency} shows that for the last $\sim$10 years of the simulation, $dE_{grav}/dE_{orbit}$ has converged to a constant value of $\approx$1.4. Furthermore, one can see from the top, right panel of Figure~\ref{fig:kippenhahn} that nearly the entire envelope above the orbit of the NS is unbound ($v/v_{esc}>1$). Therefore, the NS's inspiral rate in mass coordinate of CE during this phase can be interpreted as the rate at which the envelope material becomes unbound. This allows us to calculate the subsequent inspiral of the NS, after our simulations have stopped, using the equation:
\begin{equation}
    \Delta E_{grav}=\left(\frac{dE_{grav}}{dE_{orbit}}\right)_{f} \Delta E_{orbit},
\end{equation}
where the subscript $f$ on the $dE_{grav}/dE_{orbit}$ term refers to the value from Figure~\ref{fig:efficiency} at the end of the simulation, which we assume remains constant throughout future evolution. Replacing $\Delta E_{grav}$ and $\Delta E_{orb}$ one finds: 
\begin{eqnarray}
    & & E_{grav}(m_{c,0}) - E_{grav}(m_{c}) = \nonumber \\ 
    & & \qquad \qquad \left(\frac{dE_{grav}}{dE_{orbit}}\right)_{f}\left( \frac{Gm_{c}m_{NS}}{2r_{c}}-E_{orbit,0}\right),
\end{eqnarray}
where $r_c$ and $m_c$ are the radius and mass coordinate of the NS inside the CE, $m_{NS}$ is the mass of the NS and $G$ the gravitational constant. $E_{grav}(m_c)$ is the gravitational binding energy of the envelope, integrated from the surface down to mass coordinate $m_c$, calculated based on structure of the envelope at the last timestep of our hydrodynamic simulation. Finally, $E_{grav}(m_c,0)$ and $E_{orb,0}$ are the gravitational binding energy of the envelope down to the position of the NS and the orbital energy of the NS at the end of the hydrodynamic simulation. After solving for $r_c$, we find:
\begin{equation}
    r_{c} = \frac{1}{2}  \frac{Gm_{c}m_{NS}}{ \left(\frac{dE_{orbit}}{dE_{grav}}\right)_{f} \left[E_{grav}(m_{c,0})- E_{grav}(m_{c})\right]+E_{orbit,0}}.
    \label{eq:endCE}
\end{equation}
The solid orange line in the left panel of  Figure~\ref{fig:end_of_CE} shows the inspiral of the NS from our hydrodynamic simulation. In the right panel, we zoom in on the last stages of the CE evolution resolved by our simulation. There, the dashed orange line shows the extrapolation as predicted by Equation~\ref{eq:endCE}.

At the end of our hydrodynamic simulation, the NS is inspiraling into the CE at a steady rate of $\sim 0.003\,\rm M{_\odot}/yr$ and, equivalently, the donor is ejecting its remaining envelope at the same rate. Calculations of adiabatic mass-loss from giant stars \citep{DT2010,2010ApJ...717..724G, 2015ApJ...812...40G} show that when a large part of the envelope has been removed and the surface hydrogen abundance drops below a critical limit, the reaction of the envelope's radius to mass-loss (i.e. $\zeta_{ad}=\left(d \ln R/d \ln M\right)_{adiabatic}$) changes suddenly and the envelope contracts quickly, leading to the detachment of the binary.  

We repeat such a calculation for our initial model of the 12\,M$_\odot$ and 550\,R$_{\odot}$ giant star, rapidly removing mass by hand at four different rates. The cyan, light blue, purple, and magenta lines in Figure~\ref{fig:end_of_CE} show the radius evolution, from right to left, of the giant star as the star's mass decreases at each of these mass loss rates. All simulations show that after an initial phase of gradual expansion, once the star's mass becomes small enough (with the exact mass dependent on the specific mass-loss rate chosen), the radius sharply contracts. Given that at the end of the hydrodynamic simulation the inferred envelope mass-loss rate is $\sim 0.003\,\rm M{_\odot}/yr$, we expect that the remaining envelope will recede and the binary will detach somewhere between the crossings of the orange, dashed line with the mass-radius relations for mass-loss rates of  $0.001\,\rm M{_\odot}/yr$ (magenta line) and $0.005\,\rm M{_\odot}/yr$ (blue line). Therefore we conclude that the NS will have inspiraled for another $\sim$0.075-0.225\,M$_\odot$, to an orbital separation of $\sim$4.5-8.0\,R$_\odot$.
The duration of the remaining inspiral is expected to be $\sim25-75\,\rm yr$.

As soon as rapid mass-loss stops, the envelope is expected to re-expand on a thermal timescale to a giant star structure, albeit with a reduced thermal equilibrium radius, since the surface hydrogen abundance has now dropped to $\approx 0.3$. This implies that upon exiting a CE, the remaining hydrogen-rich envelope will overfill its Roche lobe in a semi-detached configuration and proceed to dynamically stable mass transfer on a nuclear timescale, as shown in recent work by \citet{2019arXiv190304995Q}. In their calculations, they find that these type of binaries drive mass-transfer rates of $\sim 10^{-5}\,\rm M_{\odot}\,yr^{-1}$, which in our case would translate to a mass-transfer duration of $\sim 10^5\,\rm yr$. The super-Eddington mass-transfer rate is expected to lead to a highly non-conservative mass transfer which will further shrink the orbit. Assuming that the envelope will be removed down to a surface hydrogen abundance of $\approx$10\% ($\approx$1\%) before the remaining core permanently contracts to radii characteristic of naked helium stars ($\approx$1\,R$_\odot$), one can use Equation 8 in \citet{2017A&A...597A..12S} to calculate the final orbital separation of the binary. Assuming fully non-conservative mass-transfer, this leads to final orbital separations of $\approx$3.5-5.7\,R$_\odot$ ($\approx$3.3-5.3\,R$_\odot$). 

\begin{figure}
    \centering
    \includegraphics[width=0.99\columnwidth]{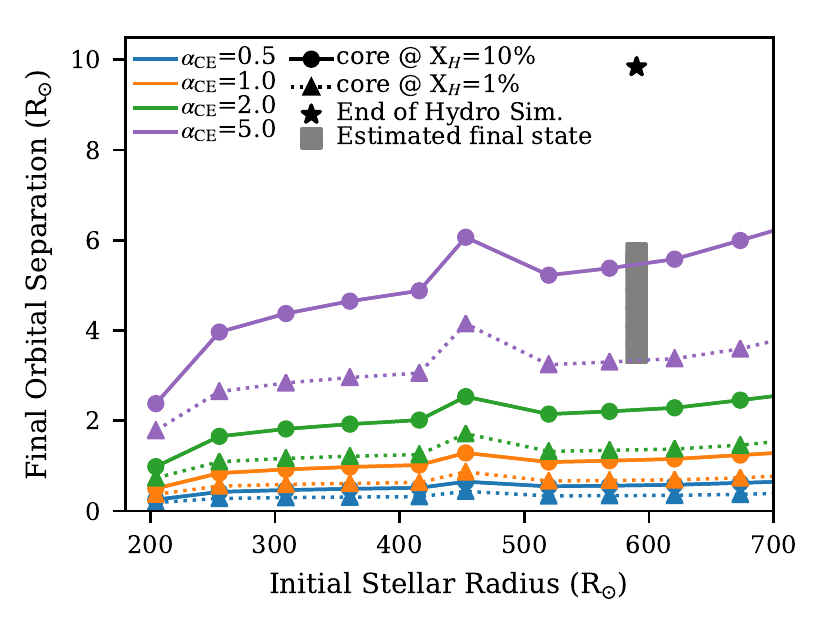}
    \caption{Comparison of the predicted final binary separation from our simulation to the application of the traditional $a_{CE}$ prescription. Colored lines show the final separation of the binary after the CE as a function of the donor's radius at the the onset of the CE, for different values of $a_{CE}$ (0.5 - blue; 1.0 - orange; 2.0 - green; 5.0 - purple) and two different definitions of the core envelope-boundary (10\% - solid lines and 1\% dashed lines).  The estimated final binary separation from this work is denoted by the grey rectangular.}
    \label{fig:alpha_prediction}
\end{figure}

From this last estimate of the final orbital separation ($a_f\approx 3.3-5.7\,\rm R_\odot$) we calculate a value of $\alpha_{CE}$ that can be directly compared to the results of rapid binary population synthesis codes. In Figure~\ref{fig:alpha_prediction} we consider binaries consisting of a 12\,M$_\odot$ giant star and a NS that initiate a CE phase at different initial orbital separations. We calculate the final orbital separation of the binary with the traditional $\alpha_{CE}$ prescription (where $\alpha_{CE}$ is defined as the ratio between the envelope's gravitational binding energy and the change in the binary orbital energy), using different values of $\alpha_{CE}$ and two different definitions of the core-envelope boundary (1\% and 10\% hydrogen fraction). Notably, the estimated final separation from our hydrodynamic simulation and the arguments we presented earlier, correspond to an $\alpha_{CE}$ value of $\sim 5$, pointing to an extremely efficient CE ejection.

\section{Conclusions}
\label{sec:conclusions}

Motivated by the challenges faced by three dimensional simulations, we model the evolution of a high-mass binary system through a common envelope using the one dimensional hydrodynamic stellar evolution code MESA. We are able to evolve a 12 \Msun\ donor star, as a 1.4 \Msun\ neutron star spirals into it, following the evolution through the three phases of CE evolution: ``loss of co-rotation", ``plunge-in", and ``self-regulated inspiral". While the initial phase of a CE may occur on a dynamical timescale, we demonstrate that the self-regulated regime occurs on a thermal timescale, where radiative transport becomes essential.

By the end of our simulation, the majority of the envelope ($\simeq 8\Msun$) has expanded to radii larger than $10^4$\Rsun, and is formally unbound from the binary. At the same time, the NS has inspiraled from an initial orbital separation of 550 \Rsun\ down to $\simeq$10 \Rsun. By analyzing the relevant energies in the system, we find the envelope is predominantly expelled by tapping into the original thermal energy of the envelope.

We estimate the final state of the binary by extrapolating the inspiral until the point where the donor star contracts. Subsequent re-expansion will cause the remaining envelope to be removed by stable mass-transfer. Such mass transfer could drive super-Eddington accretion onto a NS, potentially forming a NS ultra-luminous X-ray binary that would last for $\gtrsim 10^5$ yr. Although the binary would still be enshrouded in an optically thick envelope at a radius of $\sim$100 AU, such a phase may still be visible in X-rays. The recently observed NS ultraluminous X-ray source NGC 7793 P13 \citep{2016ApJ...831L..14F,2017MNRAS.466L..48I} is potentially such a post-CE X-ray binary \citep{2019arXiv190304995Q}.

Using a quantitative prescription for the final inspiral phase and the potential subsequent stable mass transfer, we find that the final binary consists of a $\simeq 2.6\,\rm M_{\odot}$ helium star orbiting a NS at an orbital separation of $\simeq$3.3-5.7\,\Rsun. According to detailed calculations by \citet{2003ApJ...592..475I} and \citet{2015MNRAS.451.2123T}, this binary configuration will lead to the formation of a binary NS that will merge within a Hubble time. Most interestingly, the final post CE separation translates to a very efficient CE ejection: $\alpha_{CE}\sim 5$ for the specific initial binary configuration. Our results suggest a higher efficiency than most previous studies since our simulations find that the most bound hydrogen layers surrounding the helium core are removed non-conservatively, after the binary has detached. Although there is no reason to believe that this estimated effective $\alpha_{CE}$ should have a universal value, it is tempting to think about the implications of such an efficient CE ejection. For example, it could resolve inconsistencies between rates predicted by the CE channels for coalescing binary black holes and binary NSs \citep[e.g.][]{2018MNRAS.479.4391M} as well as the formation of black hole low-mass X-ray binaries \citep[e.g.][]{PRH2003}.

\acknowledgements

We thank Bill Paxton, Pablo Marchant, Fred Rasio, and Matteo Cantiello for useful conversations. The authors acknowledge support from the  Swiss  National  Science Foundation  grants (project numbers PP00P2 17686 and 200020-172505), the Danish National Research Foundation (DNRF132), the  Marie Sklodowska-Curie RISE grant ``ASTROSTAT'' (project number 691164), the European Research Council under the European Union's Seventh Framework Programme (FP/2007-2013)/ERC Grant Agreement n. 617001, the National Science Foundation grant AST-1517753 and a CIFAR grant from the Gravity and Extreme Universe Program and a Simons Foundation grant. This work was performed in part at the Aspen Center for Physics, which is supported by National Science Foundation grant PHY-1607611.

\bibliographystyle{aasjournal}


\begin{thebibliography}{}
\expandafter\ifx\csname natexlab\endcsname\relax\def\natexlab#1{#1}\fi
\providecommand{\url}[1]{\href{#1}{#1}}



\bibitem[Abbott et al.(2017)]{2017ApJ...848L..12A} Abbott, B.~P., Abbott, R., Abbott, T.~D., et al.\ 2017, \apjl, 848, L12 


\bibitem[{Belczynski {et~al.}(2016)Belczynski, Repetto, Holz, O'Shaughnessy,
  Bulik, Berti, Fryer, \& Dominik}]{Belczynski:2016kc}
Belczynski, K., Repetto, S., Holz, D.~E., {et~al.} 2016, The Astrophysical
  Journal, 819, 108

\bibitem[{Bodenheimer \& Taam(1984)}]{Bodenheimer:1984fs}
Bodenheimer, P., \& Taam, R. 1984, \apj, 280, 771

\bibitem[{Clayton {et~al.}(2017)Clayton, Podsiadlowski, Ivanova, \&
  Justham}]{2017MNRAS.470.1788C}
Clayton, M., Podsiadlowski, P., Ivanova, N., \& Justham, S. 2017, \mnras, 470,
  1788

\bibitem[{Deloye \& Taam(2010)}]{DT2010}
Deloye, C.~J., \& Taam, R.~E. 2010, \apjl, 719, L28

\bibitem[F{\"u}rst et al.(2016)]{2016ApJ...831L..14F} F{\"u}rst, F., Walton, D.~J., Harrison, F.~A., et al.\ 2016, \apjl, 831, L14 

\bibitem[{Ge {et~al.}(2010)Ge, Hjellming, Webbink, Chen, \&
  Han}]{2010ApJ...717..724G}
Ge, H., Hjellming, M.~S., Webbink, R.~F., Chen, X., \& Han, Z. 2010, The
  Astrophysical Journal, 717, 724

\bibitem[{Ge {et~al.}(2015)Ge, Webbink, Chen, \& Han}]{2015ApJ...812...40G}
Ge, H., Webbink, R.~F., Chen, X., \& Han, Z. 2015, The Astrophysical Journal,
  812, 40

\bibitem[{{Grichener} {et~al.}(2018){Grichener}, {Sabach}, \&
  {Soker}}]{2018MNRAS.478.1818G}
{Grichener}, A., {Sabach}, E., \& {Soker}, N. 2018, \mnras, 478, 1818

\bibitem[{{Han} {et~al.}(1995){Han}, {Podsiadlowski}, \&
  {Eggleton}}]{1995MNRAS.272..800H}
{Han}, Z., {Podsiadlowski}, P., \& {Eggleton}, P.~P. 1995, \mnras, 272, 800

\bibitem[{{Ivanova}(2002)}]{Ivanova:thesis}
{Ivanova}, N. 2002, PhD thesis, University of Oxford

\bibitem[{Ivanova {et~al.}(2003)Ivanova, Belczynski, Kalogera, Rasio, Taam}]{2003ApJ...592..475I} Ivanova N., Belczynski K., Kalogera V., Rasio F.~A., Taam R.~E., 2003, ApJ, 592, 475

\bibitem[{{Ivanova}(2018)}]{2018ApJ...858L..24I}
---. 2018, \apjl, 858, L24

\bibitem[{Ivanova \& Nandez(2016)}]{2016MNRAS.462..362I}
Ivanova, N., \& Nandez, J. L.~A. 2016, \mnras, 462, 362

\bibitem[{Ivanova {et~al.}(2013)Ivanova, Justham, Chen, De~Marco, Fryer,
  Gaburov, Ge, Glebbeek, Han, Li, Lu, Marsh, Podsiadlowski, Potter, Soker,
  Taam, Tauris, van~den Heuvel, \& Webbink}]{Ivanova:2013co}
Ivanova, N., Justham, S., Chen, X., {et~al.} 2013, Astronomy and Astrophysics
  Review, 21, 59

\bibitem[Israel et al.(2017)]{2017MNRAS.466L..48I} Israel, G.~L., Papitto, A., Esposito, P., et al.\ 2017, \mnras, 466, L48 

\bibitem[{Kolb \& Ritter(1990)}]{1990A&A...236..385K}
Kolb, U., \& Ritter, H. 1990, Astronomy and Astrophysics (ISSN 0004-6361), 236,
  385

\bibitem[{Kruckow {et~al.}(2016)Kruckow, Tauris, Langer, Szecsi, Marchant, \&
  Podsiadlowski}]{2016A&A...596A..58K}
Kruckow, M.~U., Tauris, T.~M., Langer, N., {et~al.} 2016, \aap, 596, A58

\bibitem[The LIGO Scientific Collaboration et al.(2018)]{2018arXiv181112907T} The LIGO Scientific Collaboration, the Virgo Collaboration, Abbott, B.~P., et al.\ 2018, arXiv:1811.12907 

\bibitem[MacLeod, \& Ramirez-Ruiz(2015)]{2015ApJ...803...41M} MacLeod, M., \& Ramirez-Ruiz, E.\ 2015, \apj, 803, 41

\bibitem[MacLeod, \& Ramirez-Ruiz(2015)]{2015ApJ...798L..19M} MacLeod, M., \& Ramirez-Ruiz, E.\ 2015, \apjl, 798, L19

\bibitem[Mapelli \& Giacobbo(2018)]{2018MNRAS.479.4391M} Mapelli, M., \& Giacobbo, N.\ 2018, \mnras, 479, 4391 

\bibitem[{Meyer \& Meyer-Hofmeister(1979)}]{1979A&A....78..167M}
Meyer, F., \& Meyer-Hofmeister, E. 1979, \aap, 78, 167

\bibitem[{Nandez {et~al.}(2015)Nandez, Ivanova, \&
  Lombardi}]{2015MNRAS.450L..39N}
Nandez, J. L.~A., Ivanova, N., \& Lombardi, J.~C. 2015, MNRAS Letters, 450, L39

\bibitem[{Ohlmann {et~al.}(2016)Ohlmann, R{\"o}pke, Pakmor, \&
  Springel}]{Ohlmann:2016kj}
Ohlmann, S.~T., R{\"o}pke, F.~K., Pakmor, R., \& Springel, V. 2016, The
  Astrophysical Journal, 816, L9

\bibitem[{Ostriker(1999)}]{Ostriker:1999jz}
Ostriker, E.~C. 1999, \apj, 513, 252

\bibitem[{Paczy{\'{n}}ski(1976)}]{1976IAUS...73...75P}
Paczy{\'{n}}ski, B. 1976, in Structure and Evolution of Close Binary Systems;
  Proceedings of the Symposium, 75

\bibitem[{Passy {et~al.}(2012)Passy, De~Marco, Fryer, Herwig, Diehl, Oishi,
  Mac~Low, Bryan, \& Rockefeller}]{Passy:2012jo}
Passy, J.-C., De~Marco, O., Fryer, C.~L., {et~al.} 2012, \apj, 744, 52

\bibitem[{Paxton {et~al.}(2011)Paxton, Bildsten, Dotter, Herwig, Lesaffre, \&
  Timmes}]{Paxton2011}
Paxton, B., Bildsten, L., Dotter, A., {et~al.} 2011, \apjs, 192, 3

\bibitem[{Paxton {et~al.}(2013)Paxton, Cantiello, Arras, Bildsten, Brown,
  Dotter, Mankovich, Montgomery, Stello, Timmes, \&
  Townsend}]{2013ApJS..208....4P}
Paxton, B., Cantiello, M., Arras, P., {et~al.} 2013, \apjs, 208, 4

\bibitem[{Paxton {et~al.}(2015)Paxton, Marchant, Schwab, Bauer, Bildsten,
  Cantiello, Dessart, Farmer, Hu, Langer, Townsend, Townsley, \&
  Timmes}]{2015ApJS..220...15P}
Paxton, B., Marchant, P., Schwab, J., {et~al.} 2015, \apjs, 220, 15

\bibitem[{Paxton {et~al.}(2018)Paxton, Schwab, Bauer, Bildsten, Blinnikov,
  Duffell, Farmer, Goldberg, Marchant, Sorokina, Thoul, Townsend, \&
  Timmes}]{2018ApJS..234...34P}
Paxton, B., Schwab, J., Bauer, E.~B., {et~al.} 2018, \apjs, 234, 34

\bibitem[{Pejcha {et~al.}(2016{\natexlab{a}})Pejcha, Metzger, \&
  Tomida}]{2016MNRAS.455.4351P}
Pejcha, O., Metzger, B.~D., \& Tomida, K. 2016{\natexlab{a}}, \mnras, 455, 4351

\bibitem[{Pejcha {et~al.}(2016{\natexlab{b}})Pejcha, Metzger, \&
  Tomida}]{Pejcha:2016kf}
---. 2016{\natexlab{b}}, \mnras, 461, 2527

\bibitem[{Podsiadlowski(2001)}]{Podsiadlowski2001}
Podsiadlowski, P. 2001, Evolution of Binary and Multiple Star Systems; A
  Meeting in Celebration of Peter Eggleton's 60th Birthday. ASP Conference
  Series, 229, 239

\bibitem[{Podsiadlowski {et~al.}(2003)Podsiadlowski, Rappaport, \&
  Han}]{PRH2003}
Podsiadlowski, P., Rappaport, S., \& Han, Z. 2003, Monthly Notice of the Royal
  Astronomical Society, 341, 385

\bibitem[{{Quast} {et~al.}(2019){Quast}, {langer}, \&
  {Tauris}}]{2019arXiv190304995Q}
{Quast}, M., {langer}, N., \& {Tauris}, T.~M. 2019, arXiv e-prints,
  arXiv:1903.04995

\bibitem[{Rasio \& Livio(1996)}]{1996ApJ...471..366R}
Rasio, F.~A., \& Livio, M. 1996, Astrophysical Journal v.471, 471, 366

\bibitem[{Ricker \& Taam(2008)}]{Ricker:2008kc}
Ricker, P.~M., \& Taam, R.~E. 2008, \apj, 672, L41

\bibitem[{Ricker \& Taam(2012)}]{Ricker:2012gu}
---. 2012, \apj, 746, 74

\bibitem[{S{\o}rensen {et~al.}(2017)S{\o}rensen, Fragos, Steiner, Antoniou,
  Meynet, \& Dosopoulou}]{2017A&A...597A..12S}
S{\o}rensen, M., Fragos, T., Steiner, J.~F., {et~al.} 2017, \aap, 597, A12

\bibitem[{Taam(1979)}]{1979ApL....20...29T}
Taam, R. 1979, Astrophysical Letters, 20, 29

\bibitem[{Taam {et~al.}(1978)Taam, Bodenheimer, \&
  Ostriker}]{1978ApJ...222..269T}
Taam, R., Bodenheimer, P., \& Ostriker, J.~P. 1978, Astrophysical Journal, 222,
  269

\bibitem[{Taam \& Ricker(2010)}]{TR2010}
Taam, R.~E., \& Ricker, P.~M. 2010, New Astronomy Reviews, 54, 65

\bibitem[{Taam \& Sandquist(2000)}]{2000ARA&A..38..113T}
Taam, R.~E., \& Sandquist, E.~L. 2000, \araa, 38, 113

\bibitem[{Tauris \& van~den Heuvel(2006)}]{TvdH2006}
Tauris, T.~M., \& van~den Heuvel, E. P.~J. 2006, In: Compact stellar X-ray
  sources. Edited by Walter Lewin {\&} Michiel van der Klis. Cambridge
  Astrophysics Series, 623

\bibitem[Tauris et al.(2015)]{2015MNRAS.451.2123T} Tauris, T.~M., Langer, N., \& Podsiadlowski, P.\ 2015, \mnras, 451, 2123 

\bibitem[{Terman {et~al.}(1994)Terman, Taam, \&
  Hernquist}]{1994ApJ...422..729T}
Terman, J.~L., Taam, R.~E., \& Hernquist, L. 1994, Astrophysical Journal, 422,
  729

\end{thebibliography}

\end{document}